\documentstyle[epsf,epsfig,pra,aps]{revtex}

\newcommand {\beq} {\begin{equation}}
\newcommand {\eeq} {\end{equation}}
\newcommand {\bea} {\begin{eqnarray}}
\newcommand {\eea} {\end{eqnarray}}
\newcommand {\p} {\partial}
\newcommand {\sd} [1] {\tilde {#1}}
\newcommand {\D}{\mbox{d}}
\newcommand {\r} [1] {{\cal O}\left(\frac{1}{r^{#1}}\right)}

\newcommand {\R} [1] {{\cal O}\left(\frac{1}{R^{#1}}\right)}
\newcommand {\fr} [2] {\frac{#1}{r^{#2}}}
\newcommand {\fR} [2] {\frac{#1}{R^{#2}}} 
\begin{document} 
\topmargin-1.25cm

\title{Pure and gravitational radiation} 
 
\author{U. von der G\"onna and D. Kramer} 
 
\address{Institute for Theoretical Physics, FSU Jena, Max-Wien-Platz 1, 
D-07743, Germany} 
 
\maketitle 
\begin{abstract} 
The well-known treatment of asymptotically flat vacuum fields is adapted to
pure radiation fields. In this approach we find a natural normalization of the
radiation null vector. The energy 
balance at null infinity shows that the mass loss 
results from a linear superposition of the pure and 
the gravitational radiation parts. By transformation to Bondi-Sachs coordinates
the Kinnersley photon rocket is found to be the only axisymmetric Robinson-Trautman pure
radiation solution without gravitational radiation.

\vspace{5pt} 
\end{abstract} 

\section{Introduction} 

Pure radiation or null dust fields are fields of massless radiation
which is considered as the incoherent superposition of waves with
random phases and different polarizations but with the same 
propagation direction.
The radiation can arise from fields of different types, from electro-magnetic
null fields, massless scalar fields, neutrino fields or from the high frequency
limit of gravitational waves.
The energy-momentum tensor of pure radiation is
\beq
T_{ab}=\eta n_a n_b, \qquad n_a n^a=0,\qquad \eta> 0, \label{1.1}
\eeq
where $n^a$ is the tangent vector of the null congruence the radiation
propagates along and $\eta$ denotes the radiation density. The form of
(\ref{1.1}) is the result of an averaging process, i. e. the equations of the
originating fields (Maxwell equations or Weyl equation) need not to be satisfied.
However, (\ref{1.1}) can be derived from a variational principle \cite{BiKu}.

The local conservation law ${T^{ab}}_{;b}=0$ implies that the radiation
propagates along geodesics
\beq
{n^a}_{;b}n^b \propto n^a. \label{1.2}
\eeq
The energy density $\eta$ is not uniquely defined by the decomposition
(\ref{1.1}) of the energy-momentum tensor, (\ref{1.1}) is unchanged under the
following renormalization of $n^a$ and $\eta$
\beq
{\sd n}^a=f n^a \qquad \mbox{and} \qquad {\sd \eta}=\eta f^{-2}.\label{1.3}
\eeq
This annoying arbitraryness in the choice of $n^a$ even can't be avoided
by an affine parametrization of $n^a$, ${n^a}_{;b}n^b=0$, the vector $n^a$ still
might be multiplied e.~g.~by a constant factor.

Asymptotically flat pure radiation fields can describe the exterior of isolated
radiating sources. The prototype of an asymptotically flat pure radiation
field, the Vaidya solution \cite{Vaidya_53}, is often taken as the exterior
field of spherical symmetric fluid distributions. 
More general pure radiation
fields, e.~g.~the pure radiation solutions of Robinson-Trautman type,
correspondingly could be models for the exterior fields of more general isolated
sources. They are expected to contain not only pure but also
gravitational radiation.

Gravitational radiation in asymptotically flat space-times can be treated in an
elegant way by using Bondi-Sachs (BS) coordinates, and the field equations can
be solved by series expansion in terms of the luminosity parameter
\cite{Bondi_62}, \cite{Sachs_62}. In the next section we will apply this
method to the general axisymmetric asymptotically flat pure radiation field.
In order to solve the field equations step by
step in terms of power series as in the vacuum case,
the propagation vector $n^a$ of the null radiation is adapted to the luminosity
distance by fixing $n^a$ at ${\cal J}^+$.
We calculate the leading order of the radiation
density $\eta$ and the total amount of energy which is radiated away by the
null fluid. The energy balance at ${\cal J}^+$ is studied. 

In the third section the general results are applied to axisymmetric Robinson-Trautman pure
radiation solutions. By means of a transformation to BS-coordinates in terms of
power series we can fit the pure radiation field to the luminosity distance. 
The general expressions for essential quantities,
Bondi mass, news function and the leading term of the pure radiation 
energy density are given 
and we will show that the photon rocket of Kinnersley is the only solution of
this class which does not contain gravitational radiation.

{\it Notation:} All quantities related to BS-metrics are denoted by capital letters,
e.~g.~coordinates $X^a$ and metric components $G_{ab}$. The metrics have
signature $(+, +, +, -)$.

\section{Asymptotically flat pure radiation fields}

BS-coordinates $[\Theta, \Phi, R, U]$ can be introduced in the region far 
from isolated sources 
near ${\cal J}^+$ in asymptotically flat space-times. For that radially
light rays are considered. The corresponding null congruence 
is geodesic and affinely parametrized, expanding and non-rotating, such that
the
tangent vector $t^a$ is a gradient,
\beq
t_at^a=0, \qquad {t^a}_{;b}t^b=0, \qquad {t^a}_{;a}\not=0\qquad
t_a=-U_{,a}.\label{2.1}
\eeq
The function $U$ is taken as null coordinate and the intersection points of the
null geodesics with a null hypersurface $U$=const. are parametrized by the
spherical coordinates $0\le\Theta\le\pi$ and $0\le\Phi\le 2\pi$. The distance
between such an intersection point and the source is measured by the luminosity
distance $R$ which is normalized such that the closed two-surfaces $R$=const.,
$U$=const. have the surface of a sphere with the radius $R$.
Near ${\cal J}^+$, that is for sufficiently large $R$,
$R>R_0$, the line-element in BS-coordinates has 
the form
\beq
\D s^2=R^2h_{AB}(\D X^A -W^A \D U)(\D X^B- W^B \D U)
-2e^{2\beta}\D U\D R+\frac{V}{R}e^{2\beta}\D U^2,
\label{2.2}
\eeq
with 
\beq
2h_{AB}\D X^A \D X^B=(e^{2\gamma}+e^{2\delta})\D \Theta^2+4\sin \Theta
\sinh(\gamma-\delta)\D \Theta \D \Phi +\sin^2 \Theta
(e^{-2\gamma}+e^{-2\delta})\D \Phi^2. \label{2.3}
\eeq
Here $\beta, \gamma, \delta, V$ und $W^A$, $A,B\in\{1,2\}$, are functions of
the coordinates $\Theta$, $R$ and $U$, because of the axial symmetry 
they are independent of $\Phi$,
$G_{ab,\Phi}=0$. The asymptotical behaviour of the metric (\ref{2.2}) is
given by the boundary conditions. The metric functions can be expanded in
powers of $1/R$ in the coordinate range $0\le\Theta\le\pi$, $0\le\Phi\le 2\pi$,
$U_1<U<U_2$ and $R>R_0$, 
\beq
\lim_{R \rightarrow \infty}\frac{V}{R}=-1 \qquad \mbox{and} \qquad
\lim_{R \rightarrow \infty}RW^A=\lim_{R \rightarrow \infty}\beta=\lim_{R
\rightarrow \infty}\gamma=\lim_{R \rightarrow \infty}\delta=0\label{2.4}
\eeq
hold. The line element (\ref{2.2}) and the boundary conditions (\ref{2.4}) are
conserved under transformations of the Bondi-Metzner-Sachs (BMS) group 
\cite{Bondi_62}, \cite{Sachs_62}.

\begin{figure} [h]
\begin{center}
\begin{picture}(180,165)(0,0)
\put(0,0){\epsfig{figure=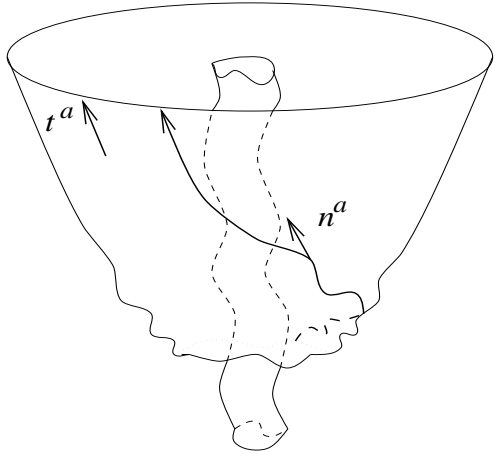,height=60mm,width=65mm}}
\end{picture} 
\end{center}
\caption{ Asymptotically flat field with pure radiation
}
\label{fig1} 
\end{figure}

We are going to consider 
space-times containing outgoing pure radiation. In general, the geodesic 
null congruence the pure radiation is propagated along is not the null
congruence (\ref{2.1}) of the coordinate system. In particular, the radiation 
geodesics may have twist.
But the asymptotic behaviour of these null congruences is the same, both
end up at ${\cal J}^+$ (see figure \ref{fig1}). The null vector $t^a$ has only a radial component,
$t^a\to(0, 0, 1, 0)$ for $R\to\infty$. The tangent vector $n^a$ of the
radiation null geodesics is also dominated by its radial component, we choose the
normalization factor such that $n^a\to(0, 0, 1, 0)$ for $R\to\infty$ holds too,
\beq
n^a=t^a+\R{ }.\label{2.5}
\eeq
With this normalization the propagation of the pure radiation is fitted to the
luminosity distance and the leading term of the 
radiation density $\eta$ is fixed.

Now we will solve the field equations in terms of power series of $1/R$ as it
has been done in the vacuum case. For relevant results the ansatz
\bea
\beta&=&\fR {\beta_1}{} +\fR {\beta_2}{ 2}+\fR{\beta_3}{3}+
\fR{\beta_4}{4}+\R 5 \label{2.6a}\\
\gamma&=&\fR {\gamma_1}{} +\fR {\gamma_2}{ 2}+\fR{\gamma_3}{3}+\R 4
\label{2.6b}\\
\delta&=&\fR {\delta_1}{} +\fR {\delta_2}{ 2}+\fR{\delta_3}{3}+\R 4 
\label{2.6c}\\
V&=&-R+2 M+\fR{q}{}+\R 2\label{2.6d}\\
W^1&=&\fR{A_2}{2}+\fR{A_3}{3}+\fR{A_4}{4}+\R 5\label{2.6e}\\
W^2&=&\fR{B_2}{2}+\fR{B_3}{3}+\fR{B_4}{4}+\R 5\label{2.6f},
\eea
with
\bea
A_3&=&2Q+\gamma_{1,\Theta}\delta_1+\delta_{1,\Theta}\gamma_1+\frac{1}{2}
(\gamma_{1,\Theta}\gamma_1+\delta_{1,\Theta}\delta_1)+4\cot \Theta
\gamma_1 \delta_1 \label{2.6g}\\
B_3&=&2S+\gamma_{1,\Theta}\gamma_1-\delta_{1,\Theta}\delta_1+\cot \Theta
(\gamma_1^2- \delta_1^2). \label{2.6h}
\eea
for the metric functions is sufficient. The regularity at the axis 
demands $\gamma$, $\delta$, $\gamma_{,\Theta}$ and $\delta_{,\Theta}$ to vanish
for $\Theta=0$ and $\Theta=\pi$.
Additionally, the ansatz
\bea
n^1&=&\hspace{20.5pt}\fR{n^{11}}{}+\fR{n^{12}}{2}+\fR{n^{13}}{3}+\R
4\label{2.7a}\\
n^2&=&\hspace{20.5pt}\fR{n^{21}}{}+\fR{n^{22}}{2}+\fR{n^{23}}{3}+\R
4\label{2.7b}\\
n^3&=&1+\fR{n^{31}}{}+\fR{n^{32}}{2}+\fR{n^{33}}{3}+\R 4\label{2.7c}\\
n^4&=&\hspace{20.5pt}\fR{n^{41}}{}+\fR{n^{42}}{2}+\fR{n^{43}}{3}+\R
4\label{2.7d}
\eea 
for the contravariant components of the propagation vector $n^a=(n^1, n^2, n^3,
n^4)$ is made, which corresponds to the normalization (\ref{2.5}).
The
condition $n^an_a=0$ for $n^a$ being null is expanded in powers of $1/R$ and
the coefficients are set equal to 0. This is the way all equations are solved.
We prefer to describe the following steps rather than to present the explicit calculations
which can easily be reproduced from these expansions.
We find 
\beq 
n^{11}=n^{21}=n^{41}=0\label{2.8}
\eeq
and the $U$-component $n^4$ of $n^a$ is fixed 
\beq
n^{42}=\frac{(n^{12})^2}{2}+\sin^2\Theta 
\frac{(n^{22})^2}{2}, \;n^{43}=\ldots\label{2.9}
\eeq
One of the ten field equations $R_{ab}=\kappa_0\eta n_an_b$ 
determines the radiation density $\eta$, the nine others are to be solved.
Starting with $R_{44}$, we have
\beq
\frac{R_{44}}{{n_4}^2}=\R 2,\label{2.10}
\eeq
i.~e.~coefficients of powers of $1/R$ lower than 2 have to vanish. So from 
$R_{33}{n_3}^{-2}={\cal O}(1/R^2)$ we get
\beq
\beta_1=0,\qquad \beta_2=-\frac{\gamma_1^2+\delta_1^2}{8},\qquad \beta_3=0.
\label{2.11}
\eeq
The corresponding conditions for $R_{13}/n_1n_3$ and $R_{23}/n_1n_3$ yield
\beq
A_2=-\frac{\gamma_{1,\Theta}+\delta_{1,\Theta}}{2}-\cot
\Theta(\gamma_1 +\delta_1) \quad \mbox{and} \quad
\sin\Theta B_2=-\frac{\gamma_{1,\Theta}-\delta_{1,\Theta}}{2}-\cot
\Theta(\gamma_1 -\delta_1). \label{2.12}
\eeq 
The coefficients of $1/R$ in $R_{11}{n_1}^{-2}$ and $R_{12}/n_1n_3$ vanish for
$\gamma_{2,U}=0=\delta_{2,U}$ and with the help of the freedom in the BMS-group
we can put
\beq
\gamma_2=0=\delta_2.\label{2.13}
\eeq
From the equation $R_{44}/{n_4}^2=\kappa_0 \eta$ the leading term of the
radiation density is calculated
\beq
\kappa_0 \eta=\fR{1}{2}\left[-2M_{,U}-\gamma_{1,U}^2-\delta_{1,U}^2
-\frac{1}{2\sin\Theta}\left[\frac{1}{\sin\Theta}\left(
\sin^2\Theta[\gamma_{1}+\delta_{1}]_{,U}\right)_{,\Theta}\right]
_{,\Theta}
\right]+\R 3.\label{2.14}
\eeq
For the remaining nine quantities $R_{ab}/n_an_b$ (no summation) the
$1/R^2$-term is set equal to $\kappa_0 \eta$. 
Only eight of these nine equations are independent (the ninth equation
corresponds to the null condition for the function $n^{42}$),
they give us a system of
algebraical equations for the functions $q$, $A_4$, $B_4$, $\beta_4$ and
the $U$-derivatives $Q_{,U}$, $S_{,U}$, $\gamma_{3,U}$ and $\delta_{3,U}$. 
This result has to be interpreted in analogy to the vacuum case. The system is
determined (up to the treated order of $1/R$) by the functions $M$, $S$, $Q$, 
$\gamma_3$ and $\delta_3$, given on a hypersurface $U$=const., and by the
functions $\gamma_1$ and $\delta_1$ and the radiation density $\eta$, which 
must be given for all $U$. 
Note
that the temporal development of the mass aspect $M$ is determined not only by the
news functions $\gamma_{1,U}$ and $\delta_{1,U}$ but also by the energy density
$\eta$ of the emitted pure radiation. 

This fact becomes even clearer if the energy balance of the system at
${\cal J}^+$ is taken into consideration. The total amount ${\cal E}$ 
of emitted  pure radiation energy is calculated by integrating the density $\eta$ over the 
closed surface $R$=const., $U$=const. in the limit $R\to\infty$. With the surface
element $R^2\sin\Theta\D\Theta\D\Phi+{\cal O}(R)$ it follows
\beq
{\cal E}=\lim\limits_{R\to\infty}2\pi\int\limits_0^{\pi}\eta R^2\sin\Theta \D
\Theta.
\eeq
Introducing the notations $\langle f \rangle$ for the average of $f$
over the intervall $0\le\Theta\le\pi$ and ${\cal M}=\langle M \rangle$ for the
Bondi mass, we get from (\ref{2.14}) with the regularity properties of
$\gamma_1$ and $\delta_1$ and the convention
$\kappa_0=8\pi$  
\beq
{\cal M}_{,U}=-{\cal E}-\frac{1}{2}\langle \gamma_{1,U}^2+\delta_{1,U}^2\rangle
\label{2.15}.
\eeq
The mass loss results from a linear superposition of pure and gravitational
radiation, provided that the propagation vector of the
pure radiation is normalized such that (\ref{2.5}) holds.
The corresponding result follows for the linear momentum. So for asymptotically
flat pure radiation fields the decomposition of the energy-momentum tensor into
propagation vector $n^a$ and energy density $\eta$ is not arbitrary, the
normalization (\ref{2.5}) is the natural relation between the physical and the
geometrical fields $n^a$ and $t^a$.

\section{Robinson-Trautman pure radiation solutions}

Robinson-Trautman space-times admit a shearfree, expanding and non-twisting
congruence of null geodesics, which is a multiple eigen congruence of the Weyl
tensor. The field equations for aligned pure radiation fields of this class are
completely solved (see \cite{KSMH} \S 24.3). In the coordinates 
$[\vartheta, \varphi, r, u]$ the line element reads
\beq
\D s^2=\frac{r^2}{P^2}(\D \vartheta^2 +\sin ^2\vartheta \D \varphi^2)-2 \D u\D r
+2H\D u^2 ,\label{3.1}
\eeq
where for the axisymmetric case
\beq
P=P(\vartheta, u) \qquad \mbox{and} \qquad H=r(\ln
P)_{,u}-\frac{K}{2}+\frac{m(u)}{r} \label{3.2}
\eeq
hold. $K/r^2$ denotes the Gaussian curvature of the surfaces $r$=const.,
$u$=const.,
which we assume to be distorted spheres,
\beq K= P^2\left[(\ln P)_{,\vartheta \vartheta}
+\cot \vartheta (\ln P)_{,\vartheta}+1\right]=:\Delta \ln P.
\label{3.3}
\eeq
The vector $k^a=(0, 0, 1, 0)$ is the tangent vector of the affinely
parametrized null geodesics, if $k^a$ is chosen as propagation vector of the
pure radiation, $R_{ab}=\kappa_0 \eta k_ak_b$, the radiation density reads
\beq
\kappa_0\eta=\frac{2}{r^2}\left[-m_{,u}+3m(\ln P)_{,u}+\frac{1}{4}\Delta \Delta
\ln P -\frac{P^2}{4}\right].\label{3.4}
\eeq
If we put $P=1$ we arrive at the Vaidya solution for which the coordinates 
$[\vartheta, \varphi, r, u]$ are BS-coordinates and by comparison with the
components of the BS-metric (\ref{2.2}) and the ansatz
(\ref{2.6a})-(\ref{2.6f}) we find ${\cal M}=m$ and $\gamma_1=\delta_1=0$.

Now we are going to the general line element (\ref{3.1}) and transform it into
BS-coordinates. The procedure is similar to that presented in \cite{Rot}, the
BS-coordinates are expanded in power series in terms of $1/r$. In the
axisymmetric non-rotating case we set
\beq
\Phi=\varphi.\label{3.5}
\eeq
By the line element of the two-surfaces $r$=const., $u$=const. the ansatz $R\sim
r/P$ is suggested, such that we start with
\bea
\Theta&=&T_0+\fr{T_1}{}+\fr{T_2}{2}+\r 3 ,\label{3.6a}\\
R&=&\frac{r}{P}+R_0+\fr{R_1}{}+\r 2 ,\label{3.6b} \\
U&=&U_0+\fr{U_1}{}+\fr{U_2}{2}+\r 3, \label{3.6c}
\eea
where all coefficients depend on $\vartheta$ and $u$. 
For the determination of these coefficients we step by step make use of the
properties
\beq
G^{44}=0=G^{14}, \quad G^{34}=-1+\R 2, \quad G^{13}=\R 2, \quad  
G^{11}G^{22}=(\sin ^2\Theta R^4)^{-1}, \label{3.7}
\eeq
the transformed 
contravariant metric components 
\beq
G^{ab}=\frac{\p X^a}{\p x^i}\frac {\p X^b}{\p x^j}g^{ij} \label{3.8}
\eeq
must have. The expansion of $G^{13}$ yields
\beq
G^{13}=-\frac{T_{0,u}}{P}+\r {}, \label{3.9}
\eeq
from which $T_{0,u}=0$ follows. Using the freedom in the BMS-group we can set
\beq
T_0=\vartheta\label{3.10}
\eeq
for simplicity. From 
\beq
G^{34}=-\frac{U_{0,u}}{P}+\r {} \label{3.11}
\eeq
we have
\beq
U_{0,u}=P ,\label{3.12}
\eeq
which fixes the $u$-dependence of $U_0$, the remaining freedom in the
$\vartheta$-dependance also corresponds to a transformation in the BMS-group.
All higher coefficients can now be expressed in terms of $T_0=\vartheta$ and
$U_0$. The condition $G^{44}=0$ yields
\beq
U_1=-\frac{1}{2}PU_{0,\vartheta}^2 \qquad \mbox{and} \qquad U_2=\frac{1}{2}
P^2U_{0,\vartheta}^2U_{0,\vartheta \vartheta} \label{3.13}
\eeq
and from $G^{14}=0$ we get
\beq
T_1=-PU_{0,\vartheta} \qquad \mbox{and} \qquad T_2=
P^2U_{0,\vartheta}U_{0,\vartheta \vartheta} .\label{3.14}
\eeq
Finally the coefficients in the expansion of $R$ are to be calculated from 
$G^{11}G^{22}=(\sin ^2\Theta R^4)^{-1}$, the leading term of that condition
confirms the ansatz $R\sim r/P$ and the higher terms give
\beq
R_0=\frac{1}{2}[U_{0,\vartheta \vartheta}+\cot \vartheta  U_{0,\vartheta}]
\label{3.15}
\eeq
and
\beq
R_1=\frac{U_{0,\vartheta}^2}{4}[P_{,\vartheta \vartheta}-\cot \vartheta
P_{,\vartheta}]-\frac{P}{8}[U_{0,\vartheta \vartheta}+\cot \vartheta
U_{0,\vartheta}]^2+\frac{P U_{0,\vartheta}}{2}[
U_{0,\vartheta \vartheta \vartheta}-\cot \vartheta
U_{0,\vartheta}] .\label{3.16}
\eeq
Thus the coefficients in the ansatz (\ref{3.6a})-(\ref{3.6c}) are determined,
terms of higher order can be calculated in the same manner by a continued 
expansion of the relevant conditions. 

The ansatz (\ref{3.6a})-(\ref{3.6c}) is
sufficient to extract characteristic quantities from the metric components.
From (\ref{2.2}) and (\ref{2.6d}) we know the component $G^{33}$ and its series
expansion in terms of $1/R$ which can be converted into a power series of $1/r$
by means of (\ref{3.6b})
\beq
G^{33}=-\frac{V}{R}e^{-2\beta}=1-2\fR{M}{}+\R 2=1-2\frac{MP}{r}+\r 2. \label{3.17}
\eeq
Now the mass aspect $M$ can be read off from the $1/r$-expansion of $G^{33}$ as
a function of $\vartheta$ and $u$
\beq
 M=\frac{m}{P^3}+\frac{ U_{0,\vartheta}^2B_{,u}}{4P}-
\frac{U_{0,\vartheta
\vartheta}B}{4}-\frac{U_{0,\vartheta}}{4}[2 B_{,\vartheta}+3\cot
\vartheta B] ,
\label{3.18}
\eeq
where $B$ is defined by
\beq
B=\frac{\sin\vartheta}{P}\left(\frac{P_{,\vartheta}}{\sin\vartheta}\right)_{,\vartheta}.
\label{3.19}
\eeq
In the non-rotating case $\gamma=\delta$ and $W^2=0$ hold for the metric
functions, such that the news is given by $\gamma_1$. As the mass aspect $M$,
the function $\gamma_1$ is calculated from the transformed metric
\beq
G^{34}=-e^{-2\beta}=-1-\frac{\gamma_1^2+\delta_1^2}{4R^2}+\R 4 
=-1-\frac{\gamma_1^2P^2}{2r^2}+\r 3 
\label{3.20}
\eeq 
as
\beq
\gamma_1^2=\frac{\sin^2\vartheta}{4}\left(\frac{U_{0,\vartheta}}{\sin\vartheta}\right)^2_{,\vartheta}.
\label{3.21}
\eeq
By differentiation with respect to $U$, which can be calculated as
\beq
\gamma_{1,u}=\gamma_{1,U}\lim \limits_{r\rightarrow \infty}\frac{\p U}{\p u}
\label{3.22}
\eeq
we get
\beq
\gamma_{1,U}^2=\frac{B^2}{4}=\frac{\sin^2\vartheta}{4P^2}
\left(\frac{P_{,\vartheta}}{\sin\vartheta}\right)^2_{,\vartheta}.
\label{3.23}
\eeq
Finally the propagation of the pure radiation is adapted to the luminosity
distance $R$, i.~e.~ the radiation null vector $n^a$ should be normalized such
that (\ref{2.5}) holds. The transformed null tetrad $k^{a'}$ 
\beq
k^{a'}=\frac{\p X^{a'}}{\p r} =(\r 2,0,\frac{1}{P}+\r 2, \r 2),\label{3.24}
\eeq
shows that $k^a$ does not fulfil the normalization condition (\ref{2.5}). 
Moreover, the
normalized propagation vector is given by
\beq
n^a=k^a\left[P+\r{ }\right] \label{3.25}
\eeq
and the leading term of the correspondingly normalized energy density of 
the pure radiation reads
\bea
\eta&=&\frac{1}{4 \pi r^2}\left[-\frac{m_{,u}}{P^2}+3m\frac{P_{,u}}{P^3}
+\frac{1}{4 P^2}\Delta \Delta
\ln P -\frac{1}{4}\right]+\r 3\nonumber \\
&=&\frac{1}{4 \pi R^2}\left[-\frac{1}{P}\left(\frac{m}{P^3}\right)_{,u}
+\frac{1}{4 P^4}\Delta \Delta
\ln P -\frac{1}{4 P^2}\right]+\R 3
.\label{3.26}
\eea

\section{The photon rocket}

Recently the 'photon rocket' solution of Kinnersley \cite{Kinn_69} has been the
subject of articles where it is proven not to contain gravitational radiation
\cite{Bonnor_94} -\cite{C_M_96}. With the results of the foregoing sections we are
now able (a) to give another proof of this fact and (b) to show that the photon
rocket is the only axisymmetric Robinson-Trautman solution with pure and without
gravitational radiation.

Kinnersley's solution is interpreted as the field of a particle emitting pure
radiation anisotropically, and accelerating because of the recoil. In the
coordinates $[\sd \vartheta, \varphi, r, u]$ the line element reads
\beq
\D s^2=r^2([\D \sd \vartheta+a\sin{\sd \vartheta}\D u]^2+\sin^2 \sd \vartheta \D \varphi^2)
-2\D u\D r-\left(1-2ar\cos\sd
\vartheta-\frac{2m}{r}\right )\D u^2,
\label{4.1}
\eeq
where the functions $m$ and $a$ depend on the retarded time $u$ and can be 
interpreted as the
mass and the acceleration of the particle, respectively. Introducing a new
coordinate $\vartheta$ by
\beq
\tan\frac{\vartheta}{2}=
e^{\displaystyle A}\tan\frac{\sd \vartheta}{2}, \qquad A=A(u),\qquad A_{,u}=a,
\label{4.2}
\eeq
the line element (\ref{4.1}) is cast into Robinson-Trautman form with the
metric function
\beq
P=\cosh A + \sinh A \cos\vartheta.
\label{4.3}
\eeq
If this function $P$ is put into the formula for the news function
(\ref{3.23}) we get $\gamma_{1,U}=0$, i.~e.~there is no gravitational
radiation. Conversely, solving the equation for vanishing news, we get
$P=b(u)+ c(u) \cos\vartheta$ which can be put into the form (\ref{4.3}) by a
coordinate transformation which preserves the form of
the Robinson-Trautman line element.

The result that the Kinnersley rocket is the only pure radiation solution
without gravitational radiation is in contradiction to \cite{Bonnor_96} where
the energy balance for a perturbed Kinnersley metric is calculated. But it
confirms the hypothesis in \cite{Darm_95} that more anisotropic pure radiation
solutions than the photon rocket would emit gravitational radiation.

\section{Conclusion}

We have been studying asymptotically flat pure radiation fields in terms of
Bondi-Sachs coordinates. In this framework we find a natural normalization for
the propagation vector of the pure radiation. The energy balance at ${\cal
J}^+$ shows a linear superposition of pure and gravitational radiation. 
The application to the Robinson-Trautman pure radiation solutions 
confirms that the Bondi-Sachs coordinates are powerful in the treatment of 
asymptotically flat fields, not only for vacuum but also for pure radiation
fields.

\end{document}